\documentclass[11pt,a4paper]{article}
\usepackage[latin1]{inputenc}
\usepackage{jcappub}
\usepackage{amsmath}
\usepackage{amsfonts}
\usepackage{amssymb}
\usepackage{graphicx}
\usepackage{epsfig}
\usepackage{color}
\usepackage{amssymb}
\usepackage{caption}

\def\beq{\begin{equation}}
\def\eeq{\end{equation}}
\def\baq{\begin{eqnarray}}
\def\eaq{\end{eqnarray}}

\newcommand{\ee}[1]{\begin{equation}#1\end{equation}}
\newcommand{\ea}[1]{\begin{align}#1\end{align}}

\newcommand{\bk}{{\bf k}}

\providecommand{\f}[2]{\frac{{#1}}{{#2}}}

\title{Despicable Dark Relics: generated by gravity with unconstrained masses}

\author[a]{Malcolm Fairbairn}
\author[b,d]{Kimmo Kainulainen}
\author[c]{Tommi Markkanen}
\author[b,d]{and  Sami Nurmi}

\affiliation[a]{Department of Physics, King's College London, Strand, London WC2R 2LS, UK}
\affiliation[b]{Department of Physics, University of Jyv\"{a}skyl\"{a},  P.O. Box 35, FI-40014 University of Jyv\"{a}skyl\"{a}, Finland}
\affiliation[c]{Department of Physics, Imperial College London,\\Blackett Laboratory, London, SW7 2AZ, United Kingdom}
\affiliation[d]{Helsinki Institute of Physics and Department of Physics, University of Helsinki, P. O. Box 64, FI-00014, Finland}

\abstract{We demonstrate the existence of a generic, efficient and purely gravitational channel producing a significant abundance of dark relics during reheating after the end of inflation. The mechanism is present for any inert scalar with the non-minimal curvature coupling $\xi R\chi^2$ and the relic production is efficient for modest values $\xi = {\cal O}(1)$. The observed dark matter abundance can be reached for a broad range of relic masses extending from  $m \sim 1 {\rm k eV}$ to $m  \sim 10^{8} {\rm GeV}$, depending on the scale of inflation and the dark sector couplings. Frustratingly, such relics escape direct, indirect and collider searches since no non-gravitational couplings to visible matter are needed.
}

\emailAdd{malcolm.fairbairn@kcl.ac.uk}
\emailAdd{kimmo.kainulainen@jyu.fi}
\emailAdd{t.markkanen@imperial.ac.uk}
\emailAdd{sami.t.nurmi@jyu.fi}

\begin{document}
\begin{flushleft}
	\hfill		  IMPERIAL/TP/2018/TM/04\\ \hfill KCL/PH-TH-2018-56
\end{flushleft}
\maketitle

\section{Introduction}
Dark matter may consist of particles which were never in chemical or kinetic equilibrium with visible matter, in contrast to thermal relics \cite{Bertone:2004pz}. The coupling of the dark sector to visible matter may be too weak to maintain equilibrium but still large enough to generate the relic abundance through out of equilibrium decays of the visible matter \cite{McDonald:2001vt}. This is commonly referred to as the freeze-in mechanism, or FIMP dark matter. See \cite{Bernal:2017kxu} for a recent review. 

Dark matter may also be completely decoupled from the visible matter and interact only gravitationally.  
A well known example is the WIMPZILLA scenario \cite{Kolb:1998ki, Chung:2001cb} where dark matter particles are produced gravitationally at the end of inflation\footnote{In \cite{Kolb:1998ki} WIMPZILLA refers to any non-thermal superheavy dark matter particle produced either gravitationally or by direct couplings, such as inflaton decays during preheating and reheating \cite{Kolb:1998ki,  Chung:1998bt}.} and must be superheavy to yield the observed relic abundance. Perturbative gravitational production may also proceed through graviton mediated scatterings after the end of inflation \cite{Garny:2015sjg, Tang:2016vch}. Moreover inflationary fuctuations of light spectators scalars, completely decoupled from the visible matter, may contribute to dark matter.  This however generates isocurvature dark matter \cite{Linde:1985yf,Nurmi:2015ema} heavily constrained by observations \cite{Ade:2015lrj}. 

Yet another, efficient and purely gravitational channel producing adiabatic dark matter was recently discovered in \cite{Markkanen:2015xuw}. Subsequently similar setups were further explored in \cite{Ema:2016hlw}. This relies on the non-minimal coupling $\xi R \chi^2$ which, under very generic conditions, is generated by radiative corrections for any energetically subdominant spectator scalar $\chi$ \cite{Chernikov:1968zm}.  During reheating, the universe is dominated by the oscillating inflaton field and the scalar curvature $R$ is an oscillatory function which periodically takes negative values. When $R$ is negative the $\chi$ particles have negative mass squared. This results in an instability and a very efficient particle production, similarly to the cases of tachyonic preheating \cite{Bassett:1997az,Tsujikawa:1999jh,Dufaux:2006ee,Bassett:1999mt, Finelli:1998bu} and vacuum instability \cite{Herranen:2015ima,Kohri:2016wof,Postma:2017hbk}. If the produced $\chi$ particles are stable and decoupled from visible matter they may constitute a natural dark matter component. 

We call these {\it despicable dark relics} for two reasons: the mechanism is very generic and in the absence of non-gravitational interactions the relics would escape all direct, indirect and collider searches of dark matter\footnote{We do not consider the effects due to possible gravitational breaking of global symmetries~\cite{Kallosh:1995hi}.}.  The particle production is efficient already for modest values of the non-minimal coupling $\xi = {\cal O}(1)$ and the relic mass window spans several orders of magnitude extending down to sub-keV scales \cite{Markkanen:2015xuw}. This is a significant difference compared to gravitationally produced WIMPZILLAs \cite{Kolb:1998ki, Chung:2001cb} which must be superheavy $m \gtrsim 10^{12}$ GeV to yield the observed dark matter abundance\footnote{Note that, like our setup, the graviton mediated scatterings discussed in \cite{Garny:2015sjg, Tang:2016vch} can also produce a significant abundance of light relics. However, this requires a very  efficient reheating whereas our setup is insensitive for the duration of reheating. It is possible that both mechanisms contribute simultaneously to the dark matter.}. If there are no isocurvature perturbations at the end of inflaton, all regions in the observable universe go through the same reheating history and acquire the same abundance of despicable relics  \cite{Markkanen:2015xuw}. Therefore, the dark matter generated through the non-minimal coupling is adiabatic as required by observations \cite{Ade:2015lrj}. 

In this work we perform a detailed investigation of the despicable dark relics.  We assume the dark sector to consist of a non-minimally coupled scalar which may have a self-interaction $\lambda \chi^4$ but no couplings to visible matter. We consider three different types of reheating stage characterised by an equation of state $w(t)$ which corresponds to inflaton oscillations in a quadratic and a quartic potential, and a kination dominated epoch. In each case we compute the final dark matter yield and find that the observed dark matter abundance can be easily obtained. In fact, the mechanism is so efficient that the spectator couplings are subject to non-trivial constraints to avoid overproduction of dark matter. 

The paper is organised as follows. In Section \ref{sec:tach} we review the analytical formalism for tachyonic particle production and in Section \ref{sec:num} we solve the equations numerically. In Section \ref{sec:ev} we study the evolution of  particle number in the presence of dark matter self-interactions and present our main results for the final dark matter yield. Finally, in Section \ref{sec:res} we present our conclusions.  

Our sign choices are (+,+,+) in the classification of \cite{Misner:1974qy}.

\section{Tachyonic particle creation at reheating}
\label{sec:tach}
  
We assume the relic scalar $\chi$ is self-interacting and non-minimally coupled to gravity but decoupled from all other fields so that its action reads 
\ee{S_\chi=-\int d^4x\,\sqrt{|g|}\bigg[\f{1}{2}(\nabla\chi)^2+\f{1}{2}m^2\chi^2+\f{\xi}{2}R\chi^2+\f{\lambda}{4}\chi^4\bigg]\,,\label{eq:act}}
where $R$ is the scalar curvature. Furthermore, we assume the spacetime dynamics during reheating are dominated by the inflaton $\phi$ and $\chi$ is an energetically subdominant spectator.     

The equation of motion for the quantised $\chi$-field is
\ee{\left(-\Box+m^2+\xi R+\lambda\hat{\chi}^2\right)\hat{\chi}=0\,. \label{eq:eom}}
The field operator $\hat{\chi}$ can be expanded in general as:
\ee{\hat{\chi}=\int \f{d^{3}{k}}{\sqrt{(2\pi )^{3}a^2}}\left[\hat{a}_\mathbf{k}^{\phantom{\dagger}}u^{\phantom{\dagger}}_{k}(\eta)+\hat{a}_{-\mathbf{k}}^\dagger u^*_k(\eta)\right]e^{i\mathbf{k\cdot\mathbf{x}}}\,\label{eq:adsol2}\, ,}
where $\eta$ is the conformal time ($ds^2=a^2(-d\eta^2+d\mathbf{x}^2$)), $\mathbf{k}$ is the co-moving momentum ($k\equiv|\mathbf{k}|$) and the creation and annihilation operators are normalized as $[\hat{a}_{\mathbf{k}}^{\phantom{\dagger}},\hat{a}_{\mathbf{k}'}^\dagger]=\delta^{(3)}(\mathbf{k}-\mathbf{k}')$. Working to one-loop accuracy
the normalized mode functions $u^{\phantom{\dagger}}_{k}(\eta)$ solve the equation 
\beq
u''_{k}(\eta)+ \bigg[\bk^2 + a^2\bigg\{ m^2+ \left(\xi-\frac{1}{6}\right) R + 3\lambda\langle\hat{\chi}^2\rangle\bigg\}\bigg] u_{k}(\eta) = 0 \,.
\label{eq:mode}\eeq
Here primes denote derivatives with respect to conformal time and the scalar curvature reads $R=6a''/a^3$. Significant excitation of modes results whenever the expression inside the square brackets of (\ref{eq:mode}) is negative. By far the most familiar example of this ``tachyonic'' instability (or spinodal decomposition) is the usual amplification of superhorizon modes during inflation, which results in the generation of an effective primordial condensate for light fields \cite{Starobinsky:1994bd}. Recently it was shown~\cite{Markkanen:2017edu} that an amplification of infrared (IR) modes is a generic feature of a light scalar field on a background characterized by the Friedmann--Lema\^itre--Robertson--Walker (FLRW) metric. 

Amplification of IR-modes during reheating may also generate a large energy density when $\xi > 1/6$. In this case the $\chi$-field is heavy during inflation so that no fluctuations are generated and the effective mass term due to self-interactions is absent until the onset of reheating. Since we assume that $\chi$ is a spectator field it has little influence on the gravitational dynamics. Hence, when the inflaton starts to oscillate during reheating so will $R$, and if the curvature term dominates over the positive mass contributions in (\ref{eq:mode}), the tachyonic instability can take place \cite{Bassett:1997az,Tsujikawa:1999jh,Dufaux:2006ee,Bassett:1999mt, Finelli:1998bu}. Indeed, the evolution of $R$ is determined by the trace of the Einstein equation:
\ee{ M_{\rm pl}^2 G_{\mu\nu}=T^\phi_{\mu\nu}=-\f{g_{\mu\nu}}{2}\bigg[\partial_\alpha\phi\partial^\alpha\phi+2V(\phi)\bigg]+\partial_\mu\phi\partial_\nu\phi\quad \Rightarrow\quad R= \f{1}{M_{\rm pl}^2}\bigg[4V(\phi)- \dot{\phi}^2\bigg]\label{eq:R(phi)}\,.}
so that we have $R<0$ whenever $|\dot \phi| > 2\sqrt{V(\phi)}$ (we use canonical kinetic terms throughout).
The precise $R$-evolution and hence the value of the generated energy density in the dark sector depends on the inflaton potential during reheating, but an oscillating or otherwise negative $R$ is a generic feature of many models.

\begin{figure}
\begin{center}
\includegraphics[width=0.9\textwidth]{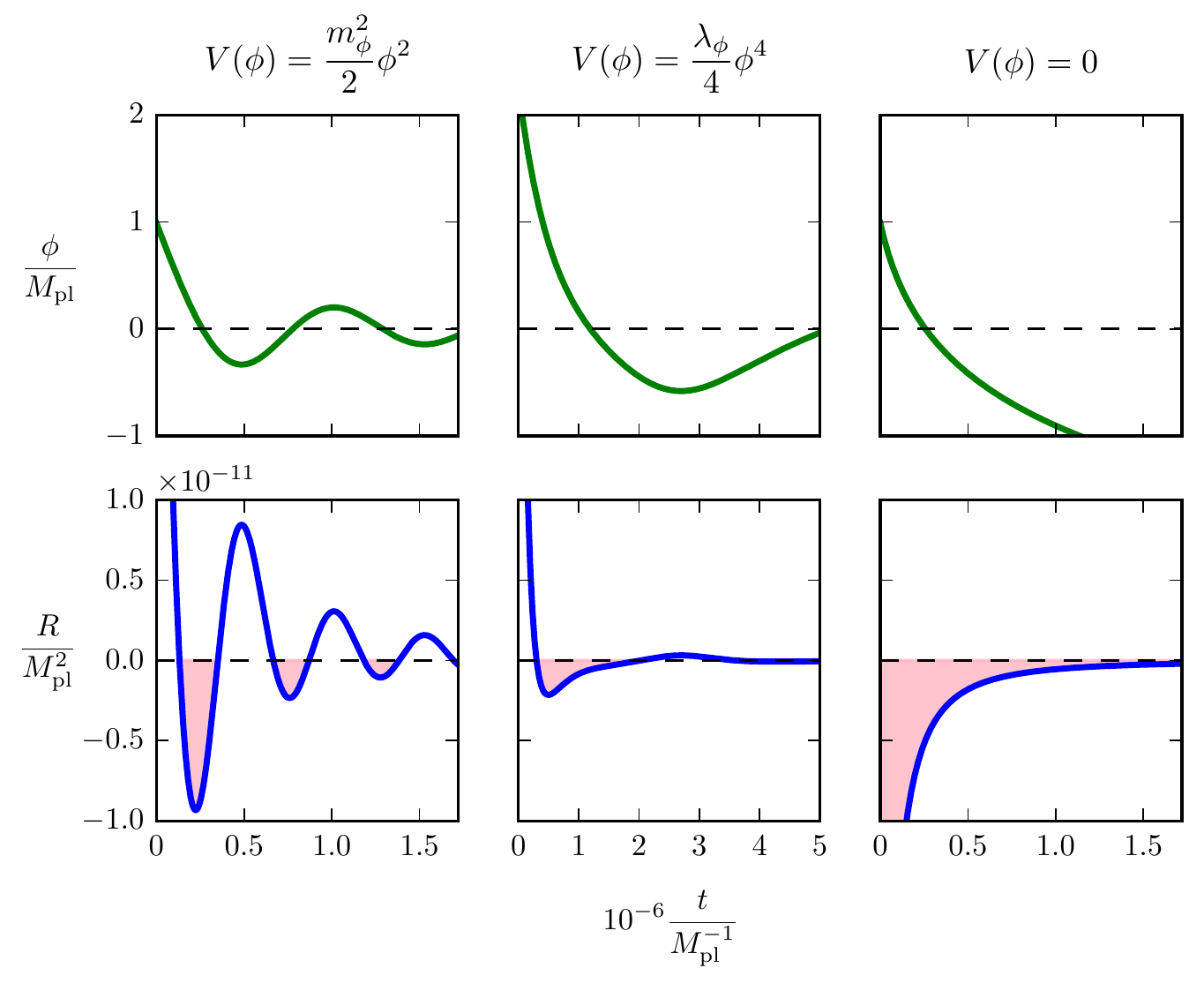}
\caption{\label{fig:oscillations} Time evolution of the inflaton $\phi$ (top) and the scalar curvature $R$ (bottom) for the three potentials (\ref{eq:dust}), (\ref{eq:radiation}) and (\ref{eq:kination}). Efficient particle production may take place in the red coloured regions with $R<0$.  We use $m_{\phi}= 1.5\times 10^{13} {\rm GeV}$ and $\lambda_{\phi}=2.8\times 10^{-12}$ and the Hubble rate at the onset of reheating is $H_{\rm inf}= 7.3\times 10^{12} {\rm GeV}$, see Section \ref{sec:res} for a detailed discussion of initial conditions.}
\end{center}
\end{figure}

We only consider the parameter region $\xi\gtrsim1$, although similar particle production during reheating is expected to take place for $\xi<0$. However, a negative non-minimal coupling leads to spontaneous symmetry breaking during inflaton complicating the analysis (see \cite{Dimopoulos:2018wfg} for an application of $\xi R$ induced symmetry breaking).

Our focus lies in three distinct choices of reheating equation of state $w(t) = p(t)/\rho(t)$ which correspond to inflaton potentials during reheating of the form 
\ea{V(\phi)&=\f{m^2_\phi}{2}\phi^2\,, &\langle w\rangle &=0\,,&
\label{eq:dust}&\\V(\phi)&=\f{\lambda_\phi}{4}\phi^4\,, &\langle w\rangle&=\f{1}{3}\,,&
\label{eq:radiation}&\\V(\phi)&=0\,,\phantom{\f{\lambda_\phi}{4}} &\langle w\rangle&={1}\,.&
&
\label{eq:kination}}
Here $\langle w\rangle$ denotes the time-averaged equation of state \cite{Turner:1983he}. Note that the form of the inflaton potential before the onset of reheating has no effect on the particle production.  
In the two first cases the inflaton oscillates in its potential and therefore also $w(t)$ and $R(t)$ oscillate around their average values.  In the last kination dominated case the equation of state stays constant $w(t) = \langle w\rangle = 1$ all the time. The time evolution of the field $\phi$ and the curvature $R$ in each of these cases is illustrated in Fig.~\ref{fig:oscillations}.  
The regions where the tachyonic amplification takes place are denoted by red. As the plots suggest, tachyonic amplification turns out to be strongly dominated by the dynamics immediately after the onset of reheating. 

Tachyonic instability cannot generate an arbitrary large energy density because of two distinct backreaction mechanisms. The first is the generation of a field dependent mass term from the self-interactions, which eventually begins to grow during preheating \cite{Bassett:1997az,Tsujikawa:1999jh,Dufaux:2006ee,Bassett:1999mt, Finelli:1998bu}. To one-loop approximation the bound for the generated variance is 
\ee{m_\chi^2 + 3\lambda\langle \hat{\chi}^2\rangle < (\xi-\frac{1}{6}) \left |R\right|\,.\label{eq:br}} 
A second backreaction effect arises when the energy density stored in the $\chi$ field can no longer be considered gravitationally irrelevant. In this case a self-consistent solution of the Einstein equation that includes the contribution of $\chi$ would be needed. But it turns out that it is in general hard to reach this threshold, as was also noticed in~\cite{Markkanen:2015xuw}.

The reheating is completed when the inflaton has fully decayed (or its energy density has been overtaken by that of the visible sector) and the radiation dominated hot big bang ensues. Details of the process depend entirely on inflaton couplings which may give rise to perturbative or non-perturbative decay channels. Here we model the inflaton decay as an instant process that takes place at $t_{\rm reh}$. We neglect any (model dependent) damping terms induced by inflaton couplings to its decay products and solve for the inflaton dynamics using the three different free potentials (\ref{eq:dust}) - (\ref{eq:kination}). The Hubble rate at the end of inflation is denoted by $H_{\rm inf}$ and at the end of reheating by $H_{\rm reh}$. We treat $H_{\rm inf}$ and $H_{\rm reh}$ as free parameters. 

After the end of reheating, the curvature induced mass vanishes $R=0$ (up to a negligible anomaly term) and the generated $\chi$ particles evolve as a decoupled dark sector. We return to the evolution and computation of the final relic abundance in Section \ref{sec:res}.

\section{A numerical approach}
\label{sec:num}

Having established that scalar curvature $R$ can generically turn negative after inflation and lead to tachyonic amplification of modes, we now make use of the adiabatic approximation to obtain a numerical solution for the generated number of particles. The adiabatic approximation is a standard approach when studying quantum fields on a curved background \cite{Birrell:1982ix} and often also used in the preheating context \cite{Dufaux:2006ee,Kofman:1997yn}.

Switching to cosmic time $t$ ($ds^2=-dt^2+a^2d\mathbf{x}^2$) and rescaling the mode functions as
\beq
v_{k}(t) \equiv\frac{u_{k}(t)}{a^{1/2}(t)}~,
\eeq 
the mode equation (\ref{eq:mode}) becomes  
\beq
\label{eq:eomN}
\ddot{v}_k(t)+\omega^2(t) v_k(t)=0~,
\eeq
where the time-dependent frequency is given by 
\baq
\label{eq:omg}
\omega^2 &=&-\f{9}{4}H^2-\f{3}{2}\dot{H}+\xi R +3\lambda\langle\hat{\chi}^2\rangle+m^2+\f{k^2}{a^2}\\
\nonumber
&=& \f{1}{M_{\rm pl}^2}\bigg[\dot{\phi}^2(3/8-\xi)+V(\phi)(4\xi-3/4)\bigg]+3\lambda\langle\hat{\chi}^2\rangle+m^2+\f{k^2}{a^2}~.
\eaq
We solve equation (\ref{eq:eomN}) numerically using the leading order adiabatic expansion where the modes are expressed as the Ansatz \cite{Dufaux:2006ee,Kofman:1997yn}
\ee{v_k(t)=\f{\alpha_\mathbf{k}}{\sqrt{2\omega}}e^{-i\int_0^t\omega}+\f{\beta_\mathbf{k}}{\sqrt{2\omega}}e^{i\int_0^t\omega}\,.\label{eq:admode}}
This is a good approximation whenever the adiabaticity conditions
\ee{\bigg\vert\f{\dot{\omega}}{\omega^2}\bigg\vert^2\lesssim 1\qquad\text{and}\qquad \bigg\vert\f{\ddot{\omega}}{\omega^3}\bigg\vert\lesssim 1\,,\label{eq:adcon}}
are satisfied. Here (\ref{eq:adcon}) are satisfied for
$\xi\gtrsim 1$,
provided that $R$ is not close to the turnover points\footnote{A careful analysis of the validity of the adiabatic expansion in the context of electroweak vacuum stability during preheating was recently performed in \cite{Postma:2017hbk}, which resulted in the bound $\xi \geq 5.5$.}. 

For $\xi\gtrsim1$ some of the modes will go through a tachyonic phase where $\omega^2<0$ and get exponentially excited. The change in the coefficients $\alpha_\mathbf{k},\beta_\mathbf{k}$ over the first tachyonic phase, and hence the number of generated particles, can be computed by matching three solutions of the form (\ref{eq:admode}) across the first two turning points surrounding the phase with $R<0$.  Up to exponentially small terms this yields \cite{Dufaux:2006ee} 
\ee{\alpha^{(1)}_\mathbf{k}=e^{X^{(1)}_\mathbf{k}}\,;\qquad\beta_\mathbf{k}^{(1)}
=-ie^{-i\Theta^{(1)}} e^{X^{(1)}_\mathbf{k}}\label{eq:bog}\,,}
where $X^{(1)}_\mathbf{k}$ is an energy integral is over the first tachyonic region $t \in [t_{1}^-,t_{1}^+]$, where $\omega_k^2<0$:
\ee{X^{(1)}_\mathbf{k}\equiv\int_{t_{1}^-}^{t_{1}^+}(-\omega^2)^{1/2}dt\,,\label{eq:X}}
and $\Theta^{(1)}$ is a phase accumulated over this epoch. The expression for $\Theta^{(1)}$ can be found in \cite{Dufaux:2006ee} but it is not relevant for our purposes.  The occupation number after the first tachyonic phase is then given by \ee{f^{(1)}({\bf k})=\vert \beta^{(1)}_\mathbf{k}\vert^2=e^{2X^{(1)}_\mathbf{k}}\,.\label{eq:n1}}
 
Particle production over subsequent tachyonic phases can be computed similarly. Neglecting quantum interference terms \cite{Dufaux:2006ee}, the occupation number at a time $t$ is given by 
\beq
\label{eq:n}
f({\bf k},t)=e^{2X_\mathbf{k}(t)}~, 
\eeq
where $X_{\mathbf{k}}(t)$ is simply the sum of integrals over all tachyonic phases 
\beq
\label{eq:X(t)}
X_{\mathbf{k}}(t) = \sum_{t_i^\pm < t} \int_{t_{i}^-}^{t_{i}^+}(-\omega_{\mathbf{k}}(t)^2)^{1/2}dt~.
\eeq
The justification for neglecting quantum phases is the expectation that in a macroscopic system classical features, such as the number of particles, persist but quantum phases rapidly decohere due to interactions with the environment, a feature which is not explicitly included in our calculation. Moreover, 
the particle production is in general dominated by the first tachyonic phase, 
which further suppresses the relevance of interference terms.

Given the occupation number (\ref{eq:n}) we can formally compute the particle number density as an integral over modes (we will omit the vacuum contributions in all integrals below~\cite{Kofman:1997yn}): 
\beq
n_{\chi}(t)= \int \frac{d^3 k}{(2\pi a(t))^3}f({\bf k},t)~.
\label{eq:n_chi}
\eeq 
However, to compute $\omega^2$ from (\ref{eq:omg}) we need the variance $\langle {\hat\chi}^2\rangle$, which itself is given by an integral over the occupation number:
\ea{\langle\hat{\chi}^2(t)\rangle&=\int 
\frac{d^3 k}{(2\pi a(t))^3}\frac{1}{\omega(t)}\bigg[\vert \beta_\mathbf{k}\vert^2+{\rm Re}\bigg(\alpha_\mathbf{k}\beta_\mathbf{k}^*e^{-i2\int_0^t\omega}\bigg)\bigg]\nonumber \\&\approx\int_0^\infty\frac{d^3 k}{(2\pi a(t))^3}\frac{1}{\omega(t)}f(\mathbf{k},t)\,.\label{eq:cond}}
(\ref{eq:X(t)}) and (\ref{eq:cond}) form a coupled set of equations for $X_k(t)$ and $\langle {\hat \chi}^2(t)\rangle$, which we have to solve numerically 
as a function of time, starting from end of inflation at $t = t_{\rm inf}$, in the FRW space governed by the inflaton:
\ea{
\begin{cases}\phantom{-(}3H^2M_{\rm pl}^2 &=\phantom{k} \f{1}{2}\dot{\phi}^2+V(\phi)\\ -(3H^2+2\dot{H})M_{\rm pl}^2 &=\phantom{k} \f{1}{2}\dot{\phi}^2-V(\phi)\end{cases}\,.\label{eq:e}}
Given $X_k(t)$ one easily finds the occupation numbers $f({\bf k},t)$ and the number density $n_\chi(t)$ from expressions (\ref{eq:n}) and (\ref{eq:n_chi}). Note the key role the variance plays in the analysis; it controls the particle production efficiency and even shuts it off if the bound (\ref{eq:br}) is violated.

The energy density of the $\chi$ field is always completely subdominant and we neglect its contribution to Friedmann equations.  Also, as already noted, the form of $V(\phi)$ during inflation has no direct relevance here; only the initial conditions at $t_{\rm inf}$ and the form of the potential during the tachyonic phase matter. 

{The variance as written in (\ref{eq:cond}) is based on the leading term in the adiabatic expansion (\ref{eq:admode}), which is valid only when (\ref{eq:adcon}) holds. In particular close to the turning points, $\omega = 0$, (\ref{eq:cond}) diverges even though physically very little particle creation is expected from this region, which is an artifact stemming from the breakdown of the adiabatic expansion. So importantly, when Im$(\omega)=0$, but $\omega$ is small, as well as when Im$(\omega)\neq0$, (\ref{eq:cond}) is not valid.  For our purposes it is important to know the variance at the point when backreaction shuts off  the tachyonic particle production as described by (\ref{eq:br}). An approximation that correctly captures this, with a well-defined expression for $\langle\hat{\chi}^2\rangle$ for all times, can be achieved by using in the denominator of (\ref{eq:cond}) the averaged frequency
\ee{\omega^2\rightarrow \langle\omega^2\rangle 
		= {\frac{9}{4}\langle w\rangle\langle H\rangle^2 } +  
		\xi\langle R\rangle+3\lambda\langle\hat{\chi}^2\rangle+m^2+\f{k^2}{\langle a\rangle^2}\,\label{eq:avR},}
where $\langle H\rangle$, $\langle R\rangle$ and $\langle a\rangle$ are given by 
\ea{\langle a\rangle&=\bigg(\f{t}{t_0}\bigg)^\f{2}{3(1+\langle w\rangle)},\quad \langle H \rangle=\f{2}{3(1+\langle w\rangle)t}\,,\nonumber\\ \langle R\rangle &= \f{4(1-3\langle w\rangle)}{3(1+\langle w\rangle)^2t^2}=3(1-3\langle w\rangle)\langle H \rangle^2\sim (1-3\langle w\rangle)\langle a\rangle^{-3(1+\langle w\rangle)}\, .\label{eq:avR2}}
We can motivate this prescription as follows: consider a situation where all modes are rendered non-tachyonic by effcient particle creation. A large mass contribution from $\lambda\langle \hat{\chi}^2\rangle$-term makes (\ref{eq:admode}) valid quickly after this, and a calculation based on $\langle \hat{\chi}^2\rangle$ from (\ref{eq:cond}) is self-consistent. In this region also the averaged frequency (\ref{eq:avR}) is correct to good accuracy and the moment when backreaction shuts off the particle creation is accurately captured. As the threshold (\ref{eq:br}) is reliably represented, the results are not expected to differ significantly from ones reached by use of more involved techniques.

\begin{figure}[!ht]
	\centering
	\includegraphics[width=0.63\linewidth]{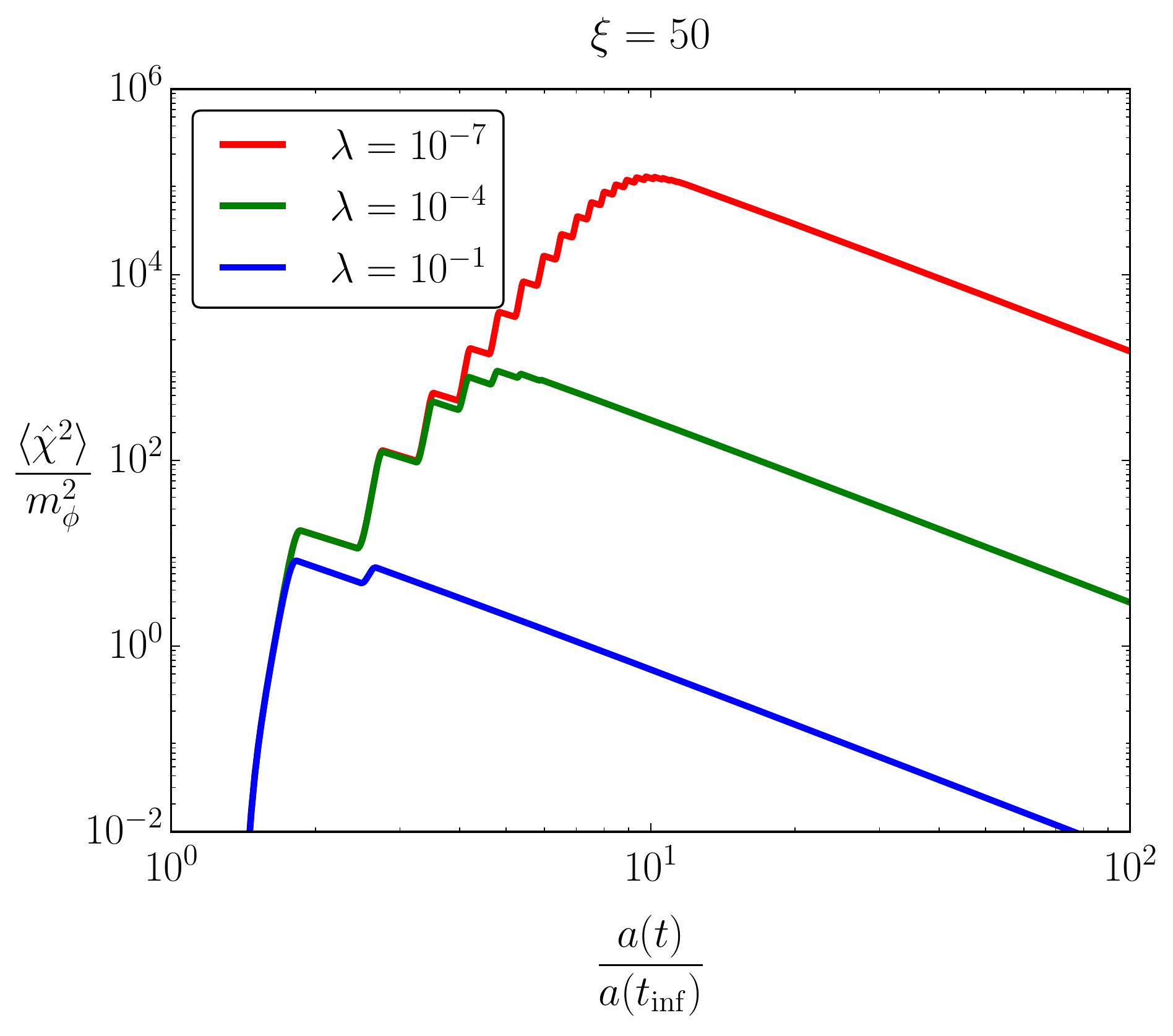}
	\caption{Evolution of the variance $\langle{\hat \chi}^2\rangle$ from the onset of reheating $t_{\rm inf}$ shown for different values of the dark matter self coupling $\lambda$. We assume quadratic inflaton potential during reheating. The maximum value of $\langle{\hat \chi}^2\rangle$ in terms of $\lambda$ can be seen to roughly scale as $\sim\langle R\rangle/\lambda\sim a(t)^{-3}/\lambda$, which is in agreement with (\ref{eq:br}).}\label{fig:lambdadamp0}
\end{figure}

Using (\ref{eq:avR}) and (\ref{eq:avR2}) we can solve (\ref{eq:X(t)}) and (\ref{eq:cond}) for $\langle\hat{\chi}^2\rangle$ and the produced DM density self-consistently as a function of time. In figure \ref{fig:lambdadamp0} we show the evolution of the variance for different values of the self-coupling of the dark matter $\lambda$ when the background is given by quadratic inflation (\ref{eq:dust}) and for $\xi=50$. The figure illustrates clearly how the field-dependent mass term $3\lambda \langle {\hat \chi}^2\rangle$ controls the tachyonic instability. For a large self coupling $\lambda = 0.1$ only one tachyonic phase contributes significantly to the variance before the threshold (\ref{eq:br}) is reached and particle production stops, after which the variance starts diluting as $a^{-2}$. Decreasing the self coupling allows for more growth of the variance, over several tachyonic phases before the threshold is reached. Note that the size of the variance correlates with the efficiency of particle production, and the above behavior as a function of $\lambda$ is only seen for a strong instability with $\xi\gg 1$.

Broadly speaking, the effect of the DM mass can be characterized in the following manner: when $m_{\chi}\ll H_{\inf}$ the tachyonic resonance at reheating will be unaffected by the DM mass and the main impact of $m_{\chi}$ is then in setting the moment beyond which the energy-density starts scaling as dust. The earlier this happens the larger the final abundance. However, when $m_\chi\gtrsim H_{\rm inf}$ the mass can prevent tachyonic particle production altogether leading to an exponential suppression of the generated energy-density. Our results do not apply in this case.

\subsection{Abundance at the end of reheating}

The reheating completes when the inflaton field has completely decayed and the universe has become radiation dominated. We denote the end of reheating by $t_{\rm reh}$. After this point the number of $\chi$ particles stays constant until a much later epoch 
when inelastic processes mediated by the self-interactions $\lambda\chi^4$ can become efficient. 

Using the conservation of the particle number we get  
\beq
a^3 \int \frac{d^3 k}{(2\pi)^3a^3}f(\mathbf{k}/a,t) = {\rm const.}~ \qquad t > t_{\rm reh},
\eeq
which implies that also the occupation number stays constant  in time 
\beq
\label{fconst}
\frac{\rm d}{{\rm d} t}f(\mathbf{k}/a(t),t) = 0 \qquad t > t_{\rm reh}~.
\eeq
Therefore, the occupation number after the end of the tachyonic phase is simply 
\beq 
f(\mathbf{k}) \equiv f(\mathbf{k},t) = f(\mathbf{k},t_{\rm reh})~, \qquad t > t_{\rm reh}~,
\eeq
where $f(\mathbf{k},t_{\rm reh})$ is computed using the numerical approach described above. 

In the radiation dominated universe $R=0$ and the curvature induced mass vanishes, up to a negligible conformal anomaly term. The particles are ultarelativistic  since we assume a small mass compared to the scale of inflation $m\ll H_{\rm inf}$ and $H_{\rm inf}$ sets the scale of tachyonically excited modes.
The number and energy densities at the onset of radiation domination are then given by 
\baq
\label{n1vsn0}
n_{\rm reh} &=& \frac{1}{a_{\rm reh}^3}\int \frac{d^3 k}{(2\pi)^3}f(\mathbf{k}) \equiv  \alpha H^3_{\rm reh}\\
\label{rho1vsrho0}
\rho_{\rm reh}&=& \frac{1}{a_{\rm reh}^4}\int \frac{d^3 k}{(2\pi)^3}k f(\mathbf{k}) \equiv  \beta H^4_{\rm reh}~,
\eaq
where the coefficients $\alpha$ and $\beta$ are determined by the numerical solution. 

For $\xi\lesssim10$ we typically we find $\alpha, \beta = {\cal O}(1)$. The strength of the mechanism is however strongly dependent on $\xi$ and $\lambda$ and for large $\xi$ and small $\lambda$ a significant density of dark relics can be produced during reheating. As an example, in Fig. \ref{grid}, we plot the $\alpha$ and $\beta$ parameters in (\ref{n1vsn0}) and (\ref{rho1vsrho0}) for the case of quadratic inflation with and $m_\phi=1.5\times 10^{13}$ GeV where the end of reheating, $a\equiv a_{\rm reh}$, occurs when $a_{\rm reh}/a_{\rm inf}=4$. 
The occupation number is highly non-thermal due to the IR nature of the tachyonic amplification. In in Fig. \ref{fofk}, we plot the the occupation number after $N_{\rm osc}$ oscillations of the inflaton field for quadratic inflation and $m_\phi=1.5\times 10^{13}$ GeV.

\begin{figure}[!ht]
	\centering
	\includegraphics[width=0.9\linewidth]{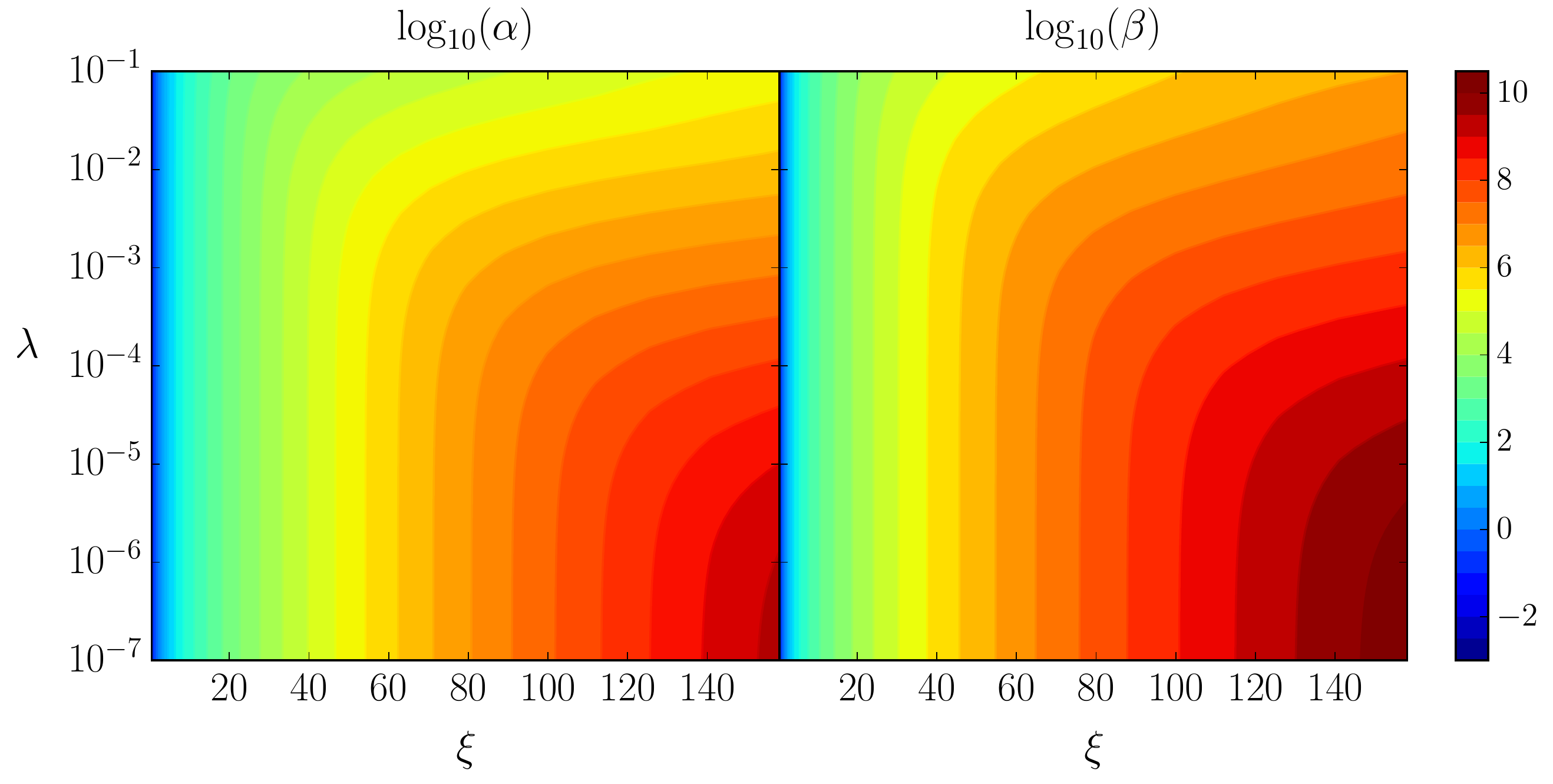}
    \caption{The $\alpha$ and $\beta$ parameters in equations (\ref{n1vsn0}) and (\ref{rho1vsrho0}) characterizing the generated particle and energy densities at the end of reheating occurring at $a_{\rm reh}/a_{\rm inf}=4$, for quadratic inflation with  $m_\phi=1.5\times10^{13}$GeV, $\lambda=1.0\times10^{-7}$ and  $\xi = 10$.\label{grid}}
\end{figure}
    \begin{figure}[!ht]
	\centering
	\includegraphics[width=0.63\linewidth]{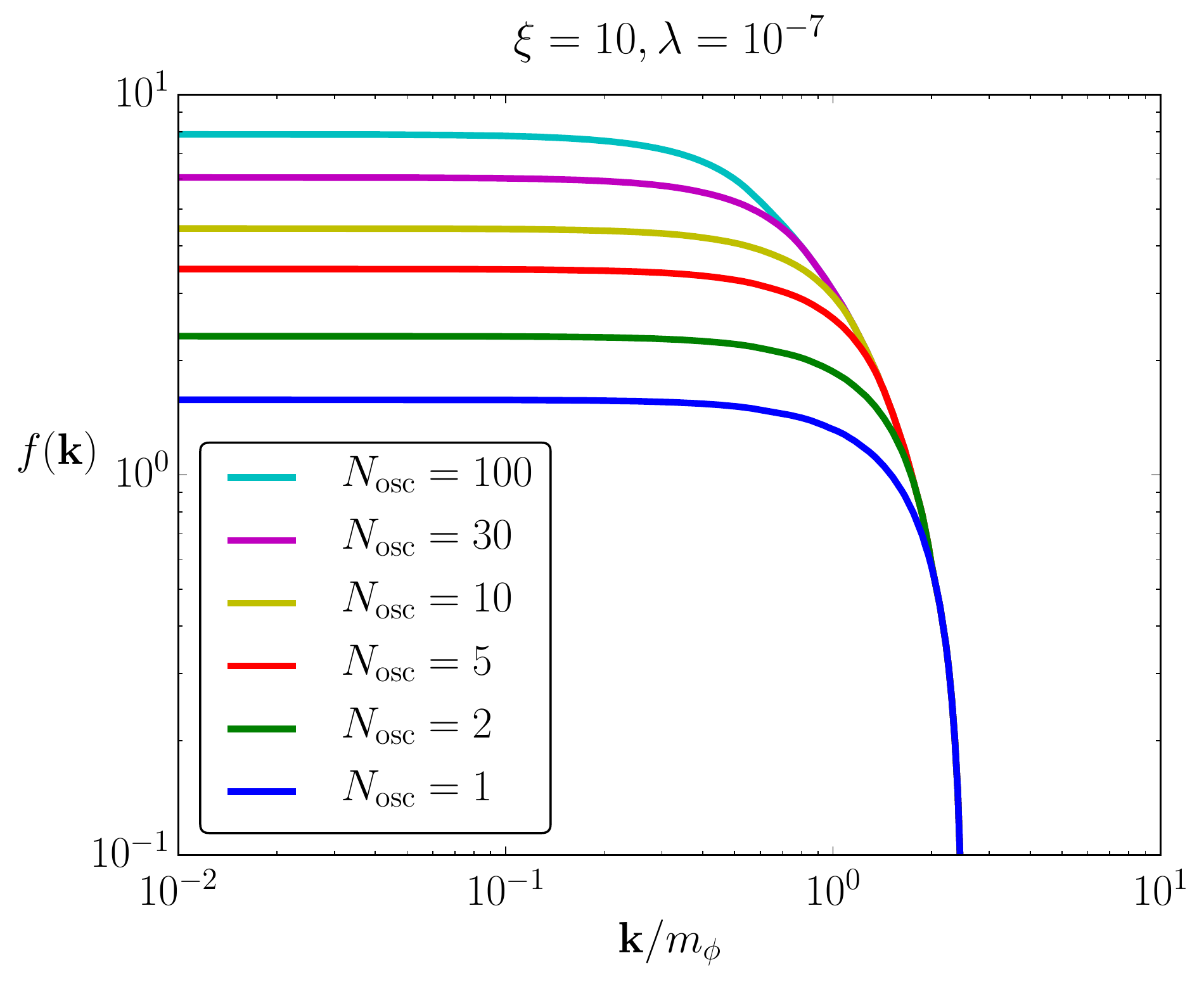}
        \caption{The occupation number $f(\mathbf{k})$ after $N_{\rm osc}$ oscillations of the inflaton in a quadratic potential with $m_\phi=1.5\times10^{13}$GeV, $\xi=10$ and $\lambda=10^{-7}$\label{fofk}. }
\end{figure}
\section{Dark thermalization and relic abundance}
\label{sec:ev}

The $\chi$ particles created by the instability never thermalize with the visible sector (to which they are only coupled gravitationally) and their distributions can evolve only through self interactions mediated by the coupling $\lambda\chi^4$. This evolution may lead to a partial or complete equilibration, and it can typically be divided into four stages: 1) non-equilibrium stage after the end of reheating $t_{\rm reh}$, 2) kinetic equilibrium established through elastic scatterings at $\Gamma_{2\leftrightarrow 2} = H_{\rm kin}$,  3) chemical equilibrium established through inelastic processes at $\Gamma_{4\rightarrow 2} = H_{\rm ch}$, and 4) dark freeze-out after the $\chi$ particles become non-relativistic. In this section we estimate the resulting dark matter abundance.  

\subsection{Stage 1: initial non-equilibrium}

Immediately after the end of reheating the $\chi$ particles are ultrarelativistic and all scattering rates are negligible. The momentum distribution retains the out-of-equilibrium form generated during the tahcyonic phase and the particle number and energy densities are given by
\baq
\label{n_1}
n_1 &=& \alpha H^3_{\rm reh}\left(\frac{a_{\rm reh}}{a}\right)^3~,\\
\label{rho_1}
\rho_{1} &=& \beta H^4_{\rm reh} \left(\frac{a_{\rm reh}}{a}\right)^4~.
\eaq
The coefficients $\alpha$ and $\beta$ are computed using the numerical methods described in Section 3. This stage continues until interactions become effective or particles become non-relativistic, whichever happens first. If $\chi$'s become non-relativistic before interactions turn on, their abundance coincides with the non-interacting case investigated in \cite{Markkanen:2015xuw}. Here we study the opposite limit, where interactions turn on while the particles are still ultrarelativistic and dark thermalisation may take place.

\subsection{Stage 2: kinetic equilibrium}

\def\lsim{\mathrel{\raise.3ex\hbox{$<$\kern-.75em\lower1ex\hbox{$\sim$}}}}

When the elastic $2\rightarrow 2$ scatterings become effective $\chi$-distribution rapidly relaxes to kinetic equilibrium. Accurate modelling of this phase would require a momentum dependent Boltzmann code and accounting for the potentially large Bose-enhancement factors, which could  induce large non-perturbative effects at infrared region~\cite{Micha:2002ey}. Such analysis is beyond the scope of this work, and since these details would not affect our final conclusions anyway, we work out the details of the kinetic equilibration in the limit of small occupation numbers (valid for $\xi \lsim 10$).  We approximate the equilibration by an instant process which happens while particles are still relativistic at 
\beq
\label{kineq}
\Gamma_{2\rightarrow 2}= H_{\rm kin}\,. 
\eeq
Assuming the Maxwell-Boltzmann form for the distributions, the thermal number and energy densities after $t_{\rm kin}$ are given by 
\beq
n_2 = \frac{e^{\mu/T} T^3}{\pi^2} 
\qquad {\rm and} \qquad
\rho_{2} =3T n_2 \,.
\label{rho_2}
\eeq
Elastic scatterings conserve the particle number and in the instant equilibration approximation their energy is also conserved over $t_{\rm kin}$. Setting $n_{1}(t_{\rm kin}) = n_2(t_{\rm kin})$ and $\rho_{1}(t_{\rm kin}) = \rho_2(t_{\rm kin})$ we get:
\beq
\label{T_2mu_2}
T_{\rm kin} =  \frac{\beta}{3\alpha}H_{\rm reh}\frac{a_{\rm reh}}{a_{\rm kin}}~
\qquad {\rm and}\qquad
e^{\mu/T}= \frac{27 \pi^2 \alpha^4}{\beta^3}~.
\eeq
After $t_{\rm kin}$ temperature scales as $T = T_{\rm kin}(a_{\rm kin}/a)$ and $\mu/T$=const. The dark sector temperature $T$ should not be confused with the temperature of the thermal bath which dominates the universe; these are in general widely different quantities.

Equation (\ref{T_2mu_2}) merely relates $T_{\rm kin}$ to the scale factor $a_{\rm kin}$. To work out $T_{\rm kin}$ in terms of physical parameters we need an explicit expression for the rate.
\beq
\label{gamma2}
\Gamma_{2\rightarrow 2} = n_2  \langle\sigma v\rangle\,,
\eeq
where the thermally averaged cross section to an arbitrary final state (of equilibrium particles) is given by~\cite{Gondolo:1990dk}:
\beq
\label{eq:GK}
\langle\sigma v\rangle = \frac{1}{8 m^4 T K_{2}^2(m/T)}\int_{s_{\rm min}}^{\infty} ds \sqrt{s}(s-4 m^2)\sigma (s)K_{1}(\sqrt{s}/T) \,,
\eeq
and $K_{1,2}$ are modified Bessel functions of the second kind. For $\chi\chi\rightarrow \chi\chi$  process $s_{\rm min} = 4 m^2$. The Bose-enhancment factors could increase this cross section significantly for large occupation numbers. Working in the ultrarelativistic limit we find
\beq
\label{gamma22}
\Gamma_{2\rightarrow 2} = n_2  \langle\sigma v\rangle  \approx   \frac{9e_{B}}{32\pi^3}\lambda^2 T e^{\mu/T}\,,
\eeq
which is valid for $m/T \lesssim 0.3$. The factor $e_{B}$ is put in by hand to model a possible Bose enhancement. Substituting (\ref{gamma22}) into (\ref{kineq}) and using that $H\propto a^{-2}$ during radiation domination yields:
\beq
\label{akin}
\frac{a_{\rm reh}}{a_{\rm kin}}={\rm min}(1,\frac{81\lambda^2\alpha^3}{32\beta^2}e_{B})~.
\eeq
Substituting (\ref{akin}) into (\ref{T_2mu_2}) gives $T_{\rm kin}$ in terms of model parameters. Particles are relativistic at this point provided that $m \ll T_{\rm kin} $, i.e. when:
\beq
\label{mass_ch}
\frac{m}{H_{\rm reh}} \ll  {\rm min}(1,\frac{27\lambda^2\alpha^2}{32\beta}e_{B})\,.
\eeq
For large masses which violate (\ref{mass_ch}), the system never reaches kinetic equilibrium and its behaviour is reduced to the non-interacting case studied in \cite{Markkanen:2015xuw}. However, for large occupation numbers the Bose factor would be of order $e_{B} \sim e^{\mu/T}$, which would in general tend to bring about an almost instantaneous kinetic equilibration during reheating.

\subsection{Stage 3: chemical equilibrium}

As is evident from (\ref{T_2mu_2}), a positive chemical potential implies an excess of particles in comparison to the full thermal equilibrium with $\mu = 0$ at the same temperature. Such excess is conserved until inelastic $2\leftrightarrow 4$ scatterings turn on and allow relaxation to chemical equilibrium. In the kinetic equilibrium, which we are now assuming holds, we may set $f(k) = e^{\mu/T}e^{-E(k)/T}$. In this case the evolution of the particle number is controlled by the equation~\footnote{Note that at this stage our calculation is no longer affected by the initially large occupation numbers: irrespective of the details of the earlier stages of the equilibration, the occupation numbers {\em are} small near chemical equilibrium, and the formulae we use in this and the following sections are valid to study the chemical equilibration and events beyond.}:
\beq
\label{lweq}
\dot{n}+3Hn
=  \langle\sigma v\rangle_{2\rightarrow4}\left(n^2 - \frac{n^4}{n_{\rm eq}^2} \right)~,
\eeq
where the average cross section $\langle\sigma v\rangle_{2\rightarrow 4}$ is given by Eq.~(\ref{eq:GK}) with $s_{\rm min} = 16m^2$. The full cross section $\sigma_{2-4}(s)$ to the four particle final state is quite involved, but in the ultrarelativistic limit $s \gg m^2$ we find a reasonably simple form:
\beq
\label{eq:xsec}
\sigma_{2-4}(s) \approx \frac{27\lambda^4}{2048 \pi^5s}\left(-{\rm Li}_2(-\frac{s}{2m^2}) + 2 \log\frac{2m^2}{s} + 4 \right) \,,
\eeq
where ${\rm Li}_2$ is the polygamma-function. Substituting this into (\ref{eq:GK}) and integrating numerically we obtain the result 
\beq
\langle\sigma v\rangle_{2\rightarrow4} \approx 2\times 10^{-5}/\sqrt{mT^3}~,
\eeq
which holds to a good approximation for $m/T \lesssim 0.1$.

Instead of solving (\ref{lweq}) numerically, we again assume that the chemical equilibration occurs in the ultrarelativistic limit and approximate it by an instant process which happens at 
\beq
\label{cheq}
\Gamma_{4\rightarrow 2}= H_{\rm ch}~. 
\eeq  
When the chemical equilibrium is established $\mu$ vanishes and the number and energy densities are given by 
\beq
\label{stage3}
n_3 =  \frac{T^3}{\pi^2}~,
\qquad {\rm and} \qquad \rho_{3} = 3T n_3~.
\eeq
Using eqs. (\ref{rho_2}) and (\ref{stage3}) and setting $\rho_{2}(t_{\rm ch}) = \rho_{3}(t_{\rm ch})$ we then find: 
\beq
\label{T_3}
T_{\rm ch} =  \left(\frac{\pi^{2}\beta}{3}\right)^{1/4} H_{\rm reh}\frac{a_{\rm reh}}{a_{\rm ch}}\,.
\eeq
After $t_{\rm ch}$ temperature again scales as $T=T_{\rm ch}(a_{\rm ch}/a)$. Note that particle number is not conserved in the chemical equilibration. Instead it decreases to a fraction determined by
\beq
\frac{n_{3}(t_{\rm ch})}{n_{2}(t_{\rm ch})}
=  \frac{1}{\sqrt{\pi}\alpha}\left(\frac{\beta}{3}\right)^{3/4} = e^{-(\mu/4T)_{\rm ch}}  < 1~.   
\eeq 

To get $T_{\rm ch}$ in terms of phycsical parameters we use (\ref{T_3}) in (\ref{cheq}), together with radiation dominance $H = H_{\rm reh}(a_{\rm reh}/a)^2$ and the explicit expression
\beq
\Gamma_{4\to 2} = n_3 \langle\sigma v\rangle_{2\rightarrow4} \approx 2\times 10^{-6} T^{3/2}/\sqrt{m}\,,
\eeq
and we find
\beq
\frac{a_{\rm reh}}{a_{\rm ch}} \approx 1.0\times 10^{-11}\beta \lambda^8 
\left(\frac{H_{\rm reh}}{m}\right)^{2}\,.
\eeq
The particles must be ultrarelativistic at this point, $m \ll T_{\rm ch}$, which holds true for coupling values in the range 
\beq
\label{mch}
\frac{m}{H_{\rm reh}} \ll \left(\frac{\lambda}{20} \right)^4\sqrt{\beta}\,.
\eeq
If the coupling is smaller,  the particles become non-relativistic and freeze-out before reaching chemical equilibrium. In this case the number of $\chi$ particles remains constant after the end of reheating and the computation of relic abundance reduces to the non-interacting case studied in \cite{Markkanen:2015xuw}.

\subsection{Stage 4: dark freeze-out}

When the dark matter particles become non-relativistic in the usual WIMP scenario, their number density becomes Boltzmann suppressed as the entropy flows out of the DM-sector. Here the situation is different: despite the initially fast $4\to 2$ interactions the number density does not get suppressed significantly, because there is no way to remove entropy from the system of $\chi$ particles which are decoupled from other matter fields (gravity mediated channels are negligible at this stage). This forces the particle number $a^3 n$ to remain nearly constant up to logarithimically small corrections. Eventually, the inelastic $4\to 2$ processes shut off due to kinematical suppression and the particle number  strictly freezes to constant. A similar process has been investigated in the context of FIMP dark matter in \cite{Carlson:1992fn}.

Precise modeling of this phase would require a momentum dependent Boltzmann code, but we can derive the approximate behaviour of the number density as follows. Assume that $n_{\rm eq}\approx (mT/2\pi)^{3/2}e^{-m/T}$ and correspondingly $s_\chi \approx (m/T + 5/2)n_{\rm eq}$. Then solve the equation $\dot s_\chi/s_\chi = - 3H$ to find the temperature $T(a)$. For $m \gg T$ the solution is $m/T \approx m/T_{\rm nr} + 3\log(a/a_{\rm nr})$, where $a_{\rm nr}$ is some initial time, and correspondingly
\begin{equation}
n_{\rm eq}(a) \approx n_{\rm eq}(T_{\rm nr})\left(\frac{a_{\rm nr}}{a}\right)^3 \left( 1 + \frac{3T_{\rm nr}}{m} \log\frac{a}{a_{\rm nr}} \right)^{-3/2} \,.
\label{eq:neq}
\end{equation}
That is, scaling deviates only logarithmically from the decoupled particle species behaviour (in fact for $m/T\lesssim 10$ deviation is even much smaller). Eventually the $4\to 2$ interactions drop out of equilibrium, not because of Boltzmann suppression, but because of a large phase space suppression of the rate in the NR-limit:
\beq
\sigma^{\rm NR}_{2\rightarrow4} \approx \frac{9}{35}\sqrt{\frac{2}{3}}\frac{\lambda^4}{512\pi^5m^2}(\frac{\sqrt{s}}{4m}-1)^{7/2}\,,
\eeq
which gives
\beq
\langle\sigma v\rangle^{\rm NR}_{2\rightarrow4} \approx \sqrt{\frac{\pi}{3}}\frac{9\lambda^4}{128\pi^2m^8}n_{\rm eq}^2\,.
\eeq
We may estimate the scale factor $a_{\rm fo}$ at which the particle number freezes to constant by setting 
\beq
\Gamma^{\rm NR}_{4\rightarrow2} = H_{\rm fo}
\eeq
with $\Gamma^{\rm NR}_{4\rightarrow2} = n_{\rm eq} \langle\sigma v\rangle^{\rm NR}_{4\rightarrow2}$, where $n_{\rm eq}$ given by (\ref{eq:neq}). Defining $a_{\rm nr}$ as the scale factor corresponding to $T_{\rm nr} = m$, we can show that $a_{\rm nr}/a_{\rm fo} \approx 0.3(m/T_{\rm ch})^{1/14}$. Using this result in (\ref{eq:neq}), we see that the effect of $4\to 2$ reactions at the onset of NR-limit is to reduce the number density by at most a factor of few in comparison to the naive decoupled species behaviour. We shall neglect such a small factor in what follows.

We can then compute number density, which in our approximation merely dilutes because of the expansion of space, using equations (\ref{stage3}) and (\ref{T_3}): 
\beq
n \approx \frac{\beta^{3/4}}{3^{3/4} \pi^{1/2}} H_{\rm reh}^3\left(\frac{a_{\rm reh}}{a}\right)^3~.
\eeq 
The corresponding energy density is given by $\rho = m n$.
Assuming usual adiabatic expansion history controlled by the standard model matter (so that $g_{*s}a^3 T_{\gamma}^3 = {\rm const.}$ where  $T_{\gamma}$ is the photon temperature), and using the values $g_{*s,0} = 3.909$ and $T_{\gamma,0} = 2.725 {\rm K}$, we find that the dark matter contribution to the energy density today is given by 
\beq
\label{omegaDM_Hreh}
\Omega_{\chi}h^2 \simeq 0.078\; \beta^{3/4}  \left(\frac{m}{{\rm GeV}}\right) \left(\frac{H_{\rm reh}}{10^{13} {\rm GeV}}\right)^{3/2}\left(\frac{100}{g_{*s,\rm reh}}\right)^{1/4} ~.
\eeq 
Here $H_{\rm reh}$ is the Hubble scale at the end of reheating. Assuming further that the average equation of state parameter $\langle w\rangle$ is constant during reheating and denoting the onset of reheating by $H_{\rm inf}$, this can be written equivalently as 
\beq
\label{omegaDM}
\Omega_{\chi}h^2 \simeq 0.078\; \beta^{3/4}  \left(\frac{m}{{\rm GeV}}\right) \left(\frac{H_{\rm inf}}{10^{13} {\rm GeV}}\right)^{3/2}\left(\frac{a_{\rm inf}}{a_{\rm reh}}\right)^{9(1+\langle w\rangle)/4}\left(\frac{100}{g_{*s,\rm reh}}\right)^{1/4} ~.
\eeq 

Eqs. (\ref{omegaDM_Hreh}) and (\ref{omegaDM}) are among the main results of this work. They give the relic abundance in the limit (\ref{mch}) where the dark sector  reaches full equilibrium before the relics become non-relativistic. The abundance is controlled by the relic mass $m$, the Hubble scale at the beginning of reheating $H_{\rm inf}$ and the $\beta$ defined in eq. (\ref{rho1vsrho0}). The $\beta$ measures  the relic energy density at the end of the tachyonic phase in units of $H_{\rm inf}$. Its dependence on $H_{\rm inf}$, $\xi$ and $\lambda$ depends on the reheating equation of state. We will investigate quadratic and quartic reheating potentials and kination domination as specific examples below. In the case of quartic potential or kination domination, the system is scale invariant and $\beta$ will not depend on $H_{\rm inf}$ at all. In the quadratic case, $\beta$ depends on $H_{\rm inf}$ through the ratio $m_{\phi}/H_{\rm inf}$ where $m_{\phi}$ is the inflaton mass term during reheating. The dependence of $\beta$ on $\xi$ and $\lambda$ in the quadratic case is depicted in Fig. \ref{grid}.

In the opposite limit, when $\lambda \lesssim 20(m/H_{\rm reh})^{1/4}\beta^{-1/8}$, particles become non-relativistic well before inelastic scatterings become efficient and the chemical equilibrium is not reached. The number of particles stays constant after the end of reheating and the relic abundance today is given by
\beq
\label{omegaDMnonrel}
\Omega_{\chi}h^2 = 0.31 {\alpha} \left(\frac{m}{\rm GeV}\right)\left(\frac{H_{\rm inf}}{10^{13} {\rm GeV}}\right)^{3/2}\left(\frac{a_{\rm inf}}{a_{\rm reh}}\right)^{9(1+w)/4}\left(\frac{100}{g_{*s,\rm reh}}\right)^{1/4} ~. 
\eeq 
This coincides with the result found in \cite{Markkanen:2015xuw} up to a factor $g_{*s,0}/g_{*s,\rm reh}$ erroneously missing from their eqs. (4.6) and (4.8). The quantity $\alpha$, defined in eq. (\ref{n1vsn0}), measures the number density of relic particles at the end of the tachyonic phase in units of $H_{\rm inf}$. The discussion on the $H_{\rm inf}$ dependence of $\beta$ above applies for $\alpha$ as well. 

\subsection{Limit on masses and couplings from dark matter self interaction}

Dark matter self interactions can be constrained by observations on dynamics of galaxies and galaxy clusters. Indeed, self interactions result in viscosity and thermal transport which can redistribute angular momentum and change the properties of DM halos.  This can be problematic since most astronomical data are well fitted by N-body simulations with collisionless dark matter. On the other hand it has been argued that dark matter self interaction can resolve some apparent astrophysical discrepancies with the simplest $\Lambda$CDM models \cite{Walker:2011zu,Bullock:2017xww}. These evidences are not conclusive however, and at least some of the observations may be explained by effects of baryons in the halos \cite{Teyssier:2012ie,Fattahi:2016nld}.

Firm upper bounds on short range self-interactions can be derived from observations of the Bullet Cluster~\cite{bullet}, an unvirialized system of two recently collided galaxy clusters, and from the galaxy NGC720, which shows evidence for an elliptical dark matter distribution.  Significant ellipticity is not compatible with a large dark matter self interaction which tends to make halos more spherical~\cite{NGC720}. These observations suggest a constraint for the self interaction cross section: $\sigma/m\le 1$cm$^2$g$^{-1} \approx 4.6\times 10^3{\rm GeV}^{-3}$.  For our simple model, where $\sigma = 9 \lambda^2/(32 \pi m^2)$ this implies a bound:
\beq
\label{eq:sigmaDMbound}
\frac{m}{\rm GeV} > 0.027 \lambda^{2/3}
\eeq
We shall see that this bound is violated in some interesting regions in parameter space.

\section{Results}
\label{sec:res}
The dark matter yield depends upon the time evolution of the equation state of the universe during reheating. This determines the duration of epochs $R<0$ where the tachyonic generation of dark matter particles takes place. 
We consider three representative examples of possible equations of state. The first two correspond to inflaton oscillations in quadratic and quartic potentials while the third represents a kination phase after inflation.  We compute the final dark matter abundance in each case.  The inflaton potential during inflation does not directly affect the dark matter yield and we leave it unspecified. Therefore, in all the three scenarios, we treat the energy scale at the onset of reheating and the inflaton initial conditions as free parameters. The same concerns the duration of the reheating epoch since it depends on the inflaton's couplings to Standard Model particles (direct or indirect) which again do not directly affect the dark matter yield. 

\subsection{Quadratic reheating potential}

We start by investigating the reheating equation of state which corresponds to inflaton oscillations in a quadratic potential $V(\phi) = m_{\phi}^2\phi^2/2$. Since we do not specify the potential during inflation, onto which the quadratic piece eventually should be glued, the precise definition of the of the end of inflation and onset of reheating $t_{\rm inf}$ is somewhat arbitrary. We choose to define $t_{\rm inf}$ as a moment when the equation of state passes $w=-1/3$ such that 
\baq
\label{Hinf_phi2}
H_{\rm inf}^2 &=& \frac{1}{3 M_{\rm pl}^2}\left(\frac{1}{2}\dot{\phi}^2
+ \frac{1}{2} m_{\phi}^2 \phi
^2 \right)~,\\
\label{winf_phi2}
- 1/3 &=& \frac{\dot{\phi}
^2-m_{\phi}^2\phi
^2}{\dot{\phi}
^2+m_{\phi}^2\phi
^2}~.
\eaq
The initial conditions are then fully determined by the initial Hubble rate $H_{\rm inf}$ and the inflaton mass $m_{\phi}$.  

We consider two different choices $H_{\rm inf}= 7.3\times 10^{12}$ GeV, $~m_{\phi}= 1.5\times 10^{13}$ GeV and $H_{\rm inf}= 7.3\times 10^{8}$ GeV, $~m_{\phi}= 1.5\times 10^{9}$ GeV.  In both cases we assume the reheating completes when $a_{\rm reh}/a_{\rm inf}=4$, where $a_{\rm reh}$ denotes the the scale factor at the end of reheating, where we assume the energy density of the inflaton field is fully converted into radiation. The expansionary period $a_{\rm reh}/a_{\rm inf}=4$ is enough time to encompass one full tachyonic part of an oscillation for both this quadratic case and the quartic case we will describe shortly. The resulting dark matter abundances computed using eqs. (\ref{omegaDM}) and (\ref{omegaDMnonrel}) are depicted in Fig. \ref{fig:quarticlam} as function of the dark matter mass $m$, self-coupling $\lambda$ and the non-minimal coupling $\xi$. 
\begin{figure}[!ht]
	\centering
	\includegraphics[width=0.75\linewidth]{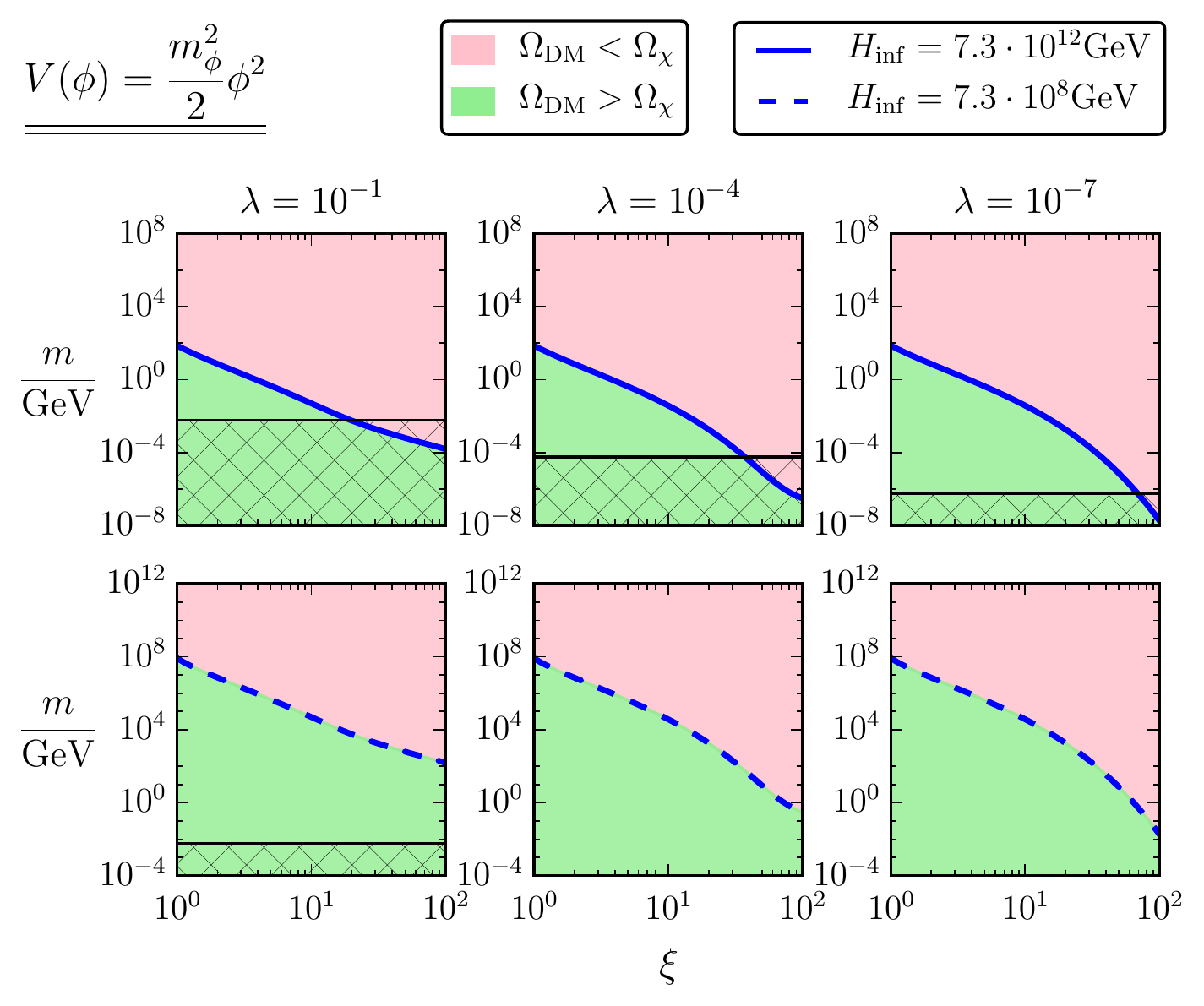}
	\caption{Relic dark matter density for the quadratic reheating potential for three different values of the dark matter self coupling $\lambda$ and two different values of the Hubble scale at the onset of reheating $H_{\rm inf}$. In all cases we assume the reheating completes when $H_{\rm reh}= H_{\rm inf}/8$ which corresponds to $a_{\rm reh}/a_{\rm inf}=4$.  In the hatched region the cross section of dark matter interactions exceeds the cluster bound  (\ref{eq:sigmaDMbound}). \label{fig:quadraticlam}}
\end{figure}

As a general trend, noted already in \cite{Markkanen:2015xuw}, we find that increasing the Hubble scale $H_{\rm inf}$ and the non-minimal coupling $\xi$ leads to stronger particle production such that a lower mass $m$ is needed to produce the observed dark matter abundance. 
Since the ratio $m_{\phi}/H_{\rm inf}$ is kept constant, the difference between the upper and lower panels is entirely due to the term $H_{\rm inf}^{3/2}$ in eqs. (\ref{omegaDM}) and (\ref{omegaDMnonrel}).
Varying $m_{\phi}/H_{\rm inf}$ would induce further dependence on $H_{\rm inf}$ since the coefficients $\beta$ and $\alpha$ in eqs. (\ref{omegaDM}) and (\ref{omegaDMnonrel}) depend on the Hubble scale through the ratio $m_{\phi}/H_{\rm inf}$. 
Increasing the self-coupling $\lambda$ decreases the dark matter abundance due to two different effects. Firstly, the tachyonic phase ends earlier due to the backreaction of generated particles which gives a growing positive contribution $\lambda\langle \chi^2\rangle$ to the effective mass squared, see Fig. \ref{fig:lambdadamp0} above and the discussion there.
Secondly, the 4-to-2 scatterings mediated through the self-coupling reduce the number of dark matter particles and therefore decrease the relic abundance. 

The effect of varying the duration of the reheating epoch, $a_{\rm reh}/a_{\rm inf}$,  is illustrated in Fig. \ref{fig:nosc}. 
\begin{figure}[!ht]
  \centering
  \includegraphics[width=0.65\linewidth]{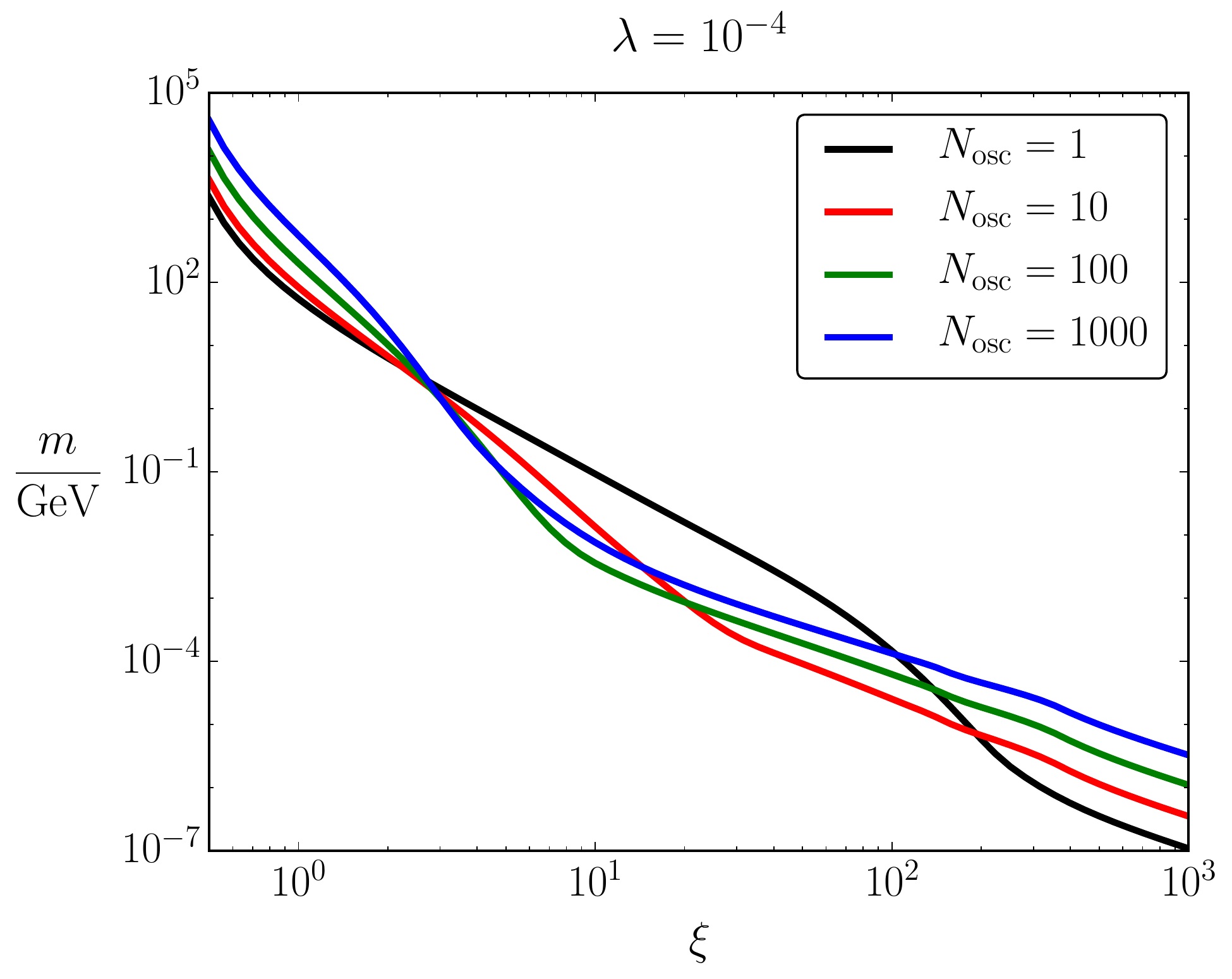}
  \caption{The figure shows how the duration of reheating affects the dark matter mass $m$ and non-minimal coupling $\xi$ which give the observed relic abundance. The duration is expressed in terms of the number of inflaton oscillations $N_{\rm osc}.$
}\label{fig:nosc}
\end{figure}
The rather complicated dependence is a result of several effects working together.  The tachyonic particle production is strongest over the first few inflaton oscillations. The subsequent oscillations give a cumulative contribution to $\langle \chi^2 \rangle $ but there is a competing opposite effect due to the redshifting of previously generated modes. For $\xi \sim 1$, the redshifting is a stronger effect and increasing the duration of the reheating epoch decreases the net production $\langle \chi^2 \rangle$.  Making $\xi$ bigger enhances the overall efficiency of the tachyonic phase and eventually turns the situation around, this is the feature seen at $\xi \sim 3$. Increasing $\xi$ further causes the tachyonic phase to shut off earlier through the backreaction from the self-interaction $\lambda$. Increasing the duration of reheating beyond this point again decreases the net yield $\langle \chi^2 \rangle$ due to the redshifting. This explains the behaviour seen in the regime $\xi \gtrsim 100$ in the figure. As a general trend, we find that the duration of reheating has the weakest effect on the relic abundance in the small $\xi$ regime. On top of this picture there are regions where the particles become non-relativistic before inelastic collisions turn on, violating the bound (\ref{mch}).

The results show that gravitational particle production during reheating can easily generate the observed dark matter abundance already for modest values of the non-minimal coupling.  Depending on $H_{\rm inf}$ and $\xi$, the dark matter mass varies over several orders of magnitude extending from $10^8$ GeV down to sub-keV scales. This is a crucial difference compared to WIMPZILLAs  \cite{Kolb:1998ki, Chung:2001cb} and related scenarios \cite{Garny:2015sjg} which only work for extremely heavy dark matter particles with masses close to the Planck scale. The purely gravitational channel discussed here is efficient over a broad mass range and, in the absence of non-gravitational couplings to visible matter, this type of dark matter could escape all direct and indirect searches - hence the name {\it despicable dark relics}. Moreover, even if the observed dark matter would be something completely different, the results place important new constraints on any SM extension with weakly coupled stable extra scalars. To avoid overproduction of despicable dark relics the couplings of the extra scalars must lie in the green regions shown in Fig.  \ref{fig:quadraticlam}.

The possibility of very low relic masses extending down to sub keV scales raises the question of Lyman $\alpha$ bounds on the scenario\footnote{We thank the anonymous Referee of the manuscript for raising up this point.} \cite{Viel:2005qj}. A detailed investigation of this topic would be an interesting subject for a future work. Here we restrict ourselves to the following qualitative argument. Assuming the despicable relics reach chemical equilibrium before becoming non-relativistic, we find from eq.~(\ref{T_3}) that their temperature $T$ is related to the photon temperature $T_{\gamma}$ parametrically through 
\beq
T \sim \beta^{1/4} \frac{T_{\gamma,{\rm reh}}}{M_{\rm pl}} T_{\gamma}~. 
\eeq
The bound the on primordial gravitational wave amplitude constrains $T_{\gamma,\rm reh}$ at least two orders below the Planck scale \cite{Akrami:2018odb}. For $\beta < 10^{9}$ we then have $T < T_{\gamma}$ and the despicable relics become non-relativistic earlier than thermal relics with the same mass. Their free streaming length is therefore smaller compared to thermal relics and, in this regime the Lyman $\alpha$ bounds are correspondingly weaker. The situation can be different for a very efficient tachyonic phase $\beta > 10^{9}$ or if the dark sector couplings are so weak that the relic states retain their initial non-equilibrium distribution, shown in Fig.~\ref{fofk}, until  when they turn non-relativistic.

\subsection{Quartic reheating potential}

Next we investigate the situation where the reheating equation of state corresponds to inflaton oscillations in a quartic potential $V = \lambda_{\phi} \phi^4/4$. In this case the mean equation averaged over inflaton oscillations is $\langle w \rangle = 1/3 $ rather than $\langle w\rangle =0$ for the quadratic case discussed above.  During the oscillations, the inflaton spends more time in the kinetic energy dominated region $\phi\approx 0$ where $R<0$ (\ref{eq:R(phi)}) and therefore we expect enhanced production of dark matter.  

We start the evolution in the same way as for the $\phi^2$ case before. We define the onset of reheating $t_{\rm inf}$ as a time when the equation of state passes $w= -1/3$ and give the corresponding Hubble scale $H_{\rm inf}$ and $\lambda_{\phi}$ as the initial data. The initial inflaton field value and its derivative are then fully determined by relations similar to eqs. (\ref{Hinf_phi2}) and \ref{winf_phi2}). To compare with the quadratic case we use the same two choice for the initial Hubble scale 
$H_{\rm inf}= 7.3\times 10^{12} {\rm GeV}$ and $ H_{\rm inf}= 7.3\times 10^{8} {\rm GeV}$. In both cases we set $\lambda_{\phi}= 2.8\times10^{-12}$ and use $a_{\rm reh}/a_{\rm inf}=4$. 

The results for the dark matter abundance are shown in Fig. \ref{fig:quarticlam}. 
\begin{figure}[!ht]
  \centering
  \includegraphics[width=0.75\linewidth]{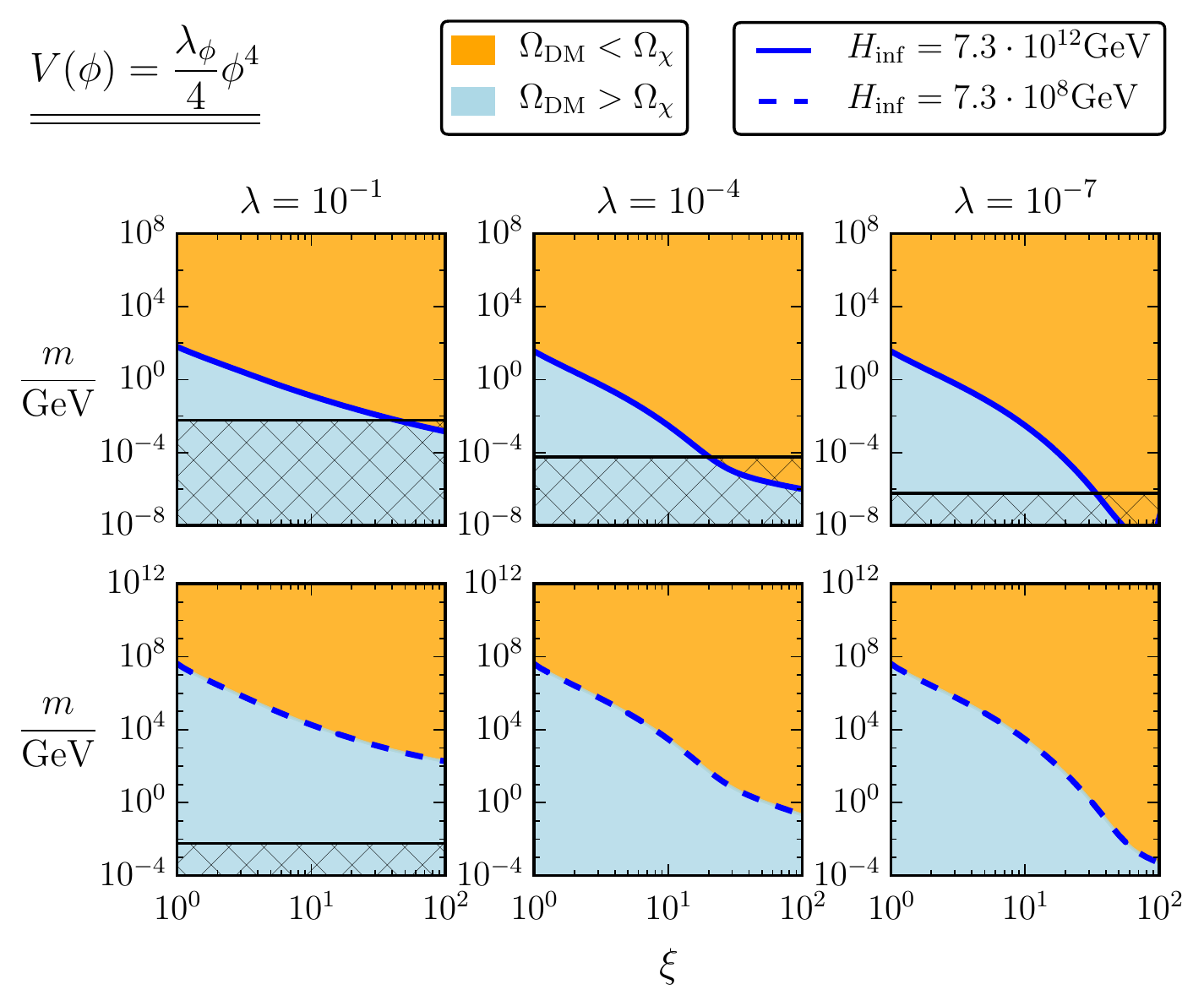}
  \caption{Relic dark matter density for the quartic reheating potential for three different values of the dark matter self coupling $\lambda$ and two different values of the Hubble scale at the onset of reheating $H_{\rm inf}$. In all cases we assume the reheating completes when $H_{\rm reh}= H_{\rm inf}/16$ which corresponds to $a_{\rm reh}/a_{\rm inf}=4$.\label{fig:quarticlam}}
\end{figure}
Comparing to Fig. \ref{fig:quadraticlam}, it is seen that the observed dark matter abundance is here obtained for smaller masses than in the quadratic case. This is due to the more prolonged tachyonic periods where $R<0$ which lead to enhanced particle production. 
As discussed above, the quantities $\beta$ and $\alpha$ in eqs. (\ref{omegaDM}) and (\ref{omegaDMnonrel}) do not depend on $H_{\rm inf}$ in this case since the quartic reheating potential is scale invariant. Therefore, the difference between the upper and lower panels is entirely due to the term $H_{\rm inf}^{3/2}$ of eqs. (\ref{omegaDM}) and (\ref{omegaDMnonrel}).
The results are relatively robust against changing the duration of the reheating: we have checked that varying $a_{\rm reh}/a_{\rm inf}$ from $1$ to $1000$ amounts to order of magnitude changes in the values $m$ and $\xi$, which correspond to  correct relic abundance. This is qualitatively similar to the behaviour in the quadratic case depicted in Fig. \ref{fig:nosc} despite the different redshifting of the background energy density.

\subsection{Kination dominated reheating}

As the last example, we study kination dominated reheating where the equation of state is $w=1$. In this case $R = - 2 \rho/M_{\rm pl}^2$ is negative all the time and the efficiency of the tahcyonic particle production is maximal. The particle production will continue constantly until it is cut off by backreaction due to the self coupling $\lambda$, or the reheating ends when the universe becomes radiation dominated $R=0$. 

The kination phase can be realised when the inflaton kinetic energy dominates over the potential \cite{Joyce:1996cp,Ferreira:1997hj}.  Here we simply assume that inflaton potential vanishes $V=0$. In this case the initial conditions are specified by the value of the Hubble rate when we start the evolution $H_{\rm inf} = \sqrt {\phi}/ (\sqrt{6} M_{\rm pl})$ (the initial field value $\phi_{\rm inf}$ is irrelevant). As before, we consider two choices $H_{\rm inf}= 7.3\times 10^{12} {\rm GeV}$ and $ H_{\rm inf}= 7.3\times 10^{8} {\rm GeV}$ and set $a_{\rm reh}/a_{\rm inf}=4$. 
\begin{figure}[!ht]
  \centering
  \includegraphics[width=0.75\linewidth]{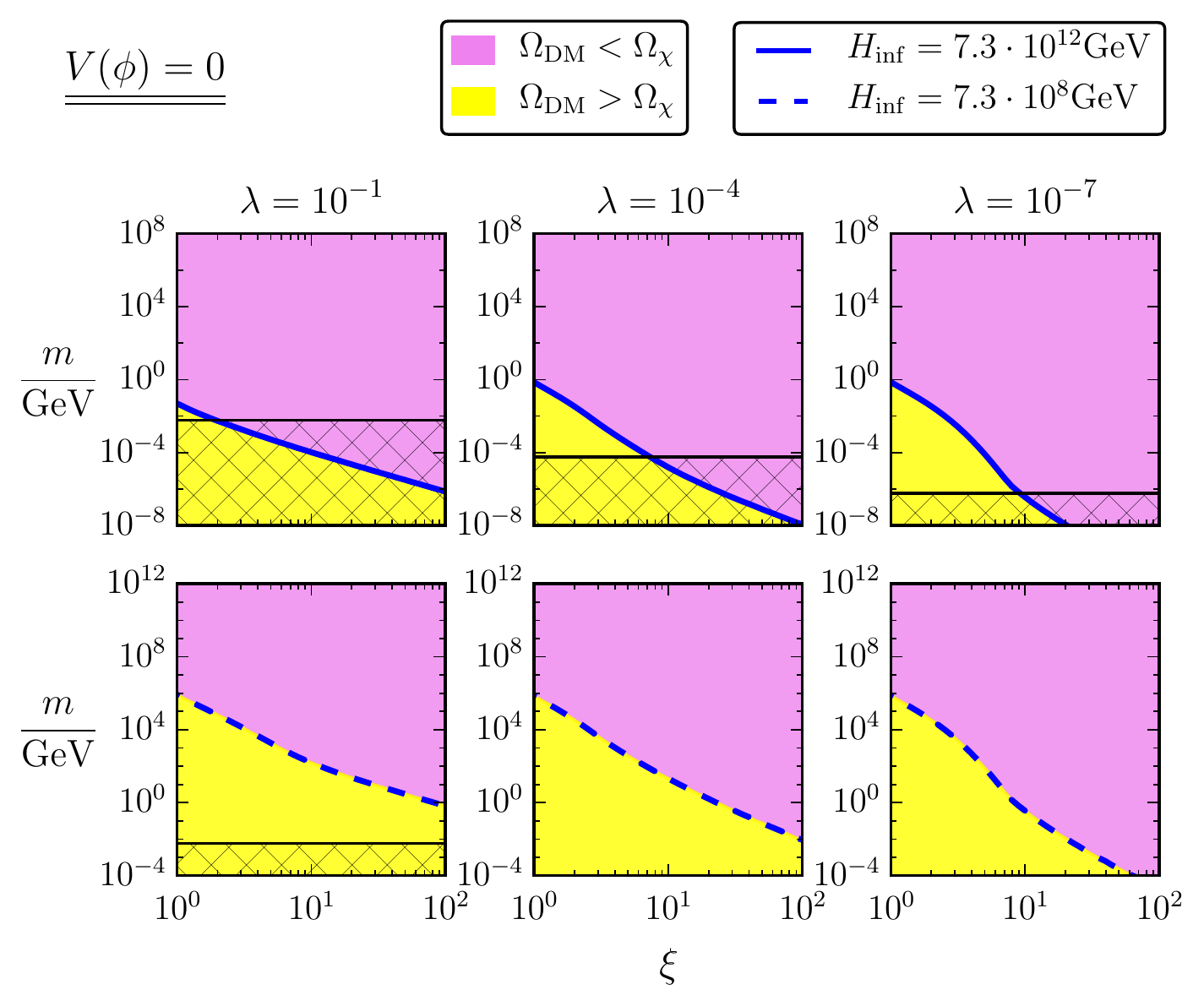}
  \caption{Relic density for the kination dominated reheating period. In all cases we assume the reheating completes when $H_{\rm reh}= H_{\rm inf}/64$ which corresponds to $a_{\rm reh}/a_{\rm inf}=4$.\label{fig:kinationlam}}
\end{figure}
The results for the dark matter abundance are shown in Fig. \ref{fig:kinationlam}. Comparing to the previous two examples, we see that a kination dominated reheating leads to the most efficent dark matter production as expected. Again the quantities $\beta$ and $\alpha$ in eqs. (\ref{omegaDM}) and (\ref{omegaDMnonrel}) do not depend on $H_{\rm inf}$ as the kination dominated reheating phase is scale invariant. The difference between the upper and lower panels is therefore entirely due to the term $H_{\rm inf}^{3/2}$ in eqs. (\ref{omegaDM}) and (\ref{omegaDMnonrel}).

\section{Conclusions}

In this work we have explored a mechanism, first set out in \cite{Markkanen:2015xuw}, where dark matter is produced gravitationally during reheating. We find that the mechanism is very efficient, and the observed dark matter abundance can be reached for a very broad range of relic masses. 

We have investigated three different reheating scenarios, corresponding to inflaton oscillations in quadratic and quartic potentials, and a kination epoch. In all the cases the curvature scalar $R$ evolves to negative values which may trigger a tachyonic instability for scalars with the non-minimal coupling $\xi R \chi^2$. If the produced particles are stable, they constitute adiabatic dark matter \cite{Markkanen:2015xuw}. We have focused on a dark sector consisting of a single non-minimally coupled scalar $\chi$ with a mass $m$ and a self-interaction $\lambda \chi^4$, and no couplings to any other matter fields. We have concentrated on the region $\xi \gtrsim 1$ where the scalar does not fluctuate during inflation but can experience a strong tachyonic instability during reheating. We have performed a detailed numerical analysis of the particle production for each scenario, varying the reheating scale, its duration, the non-minimal coupling $\xi$ and the self-interaction strength $\lambda$ which all affect the yield. We have followed the evolution of the relic particle number from the end of reheating until today,  accounting in particular for the possible impacts of inelastic scatterings mediated through the self-coupling $\lambda \chi^4$. 

The main results of this work are encompassed in Figures \ref{fig:quadraticlam}, \ref{fig:quarticlam} and \ref{fig:kinationlam} which show the present day abundance of gravitational relics. The observed dark matter abundance can be obtained in broad range of relic masses, extending down to keV scale, 
and for realistic dark sector couplings $\lambda$ and $\xi$. This should be contrasted to gravitationally produced WIMP-ZILLAs which must be superheavy $m\gtrsim 10^{12}$ GeV to yield the observed abundance \cite{Kolb:1998ki, Chung:2001cb}. The difference stems from the efficiency of the tachyonic particle production mechanism. For $\xi\gg 1$ the mechanism can in fact easily yield too much dark matter, which opens up a potentially interesting new way to constrain theories with stable extra scalars. 

The key features of the scenario are its genericity and that the relics can be completely decoupled from visible matter. The fact that they would thus remain undetectable in all conceivable dark matter searches, 
motivates us to introduce the name Despicable Dark Relics (DDR) for this type of dark matter.

\section*{Acknowledgments}
The research leading to these results has received funding from the European Research Council under the European Union's Horizon 2020 program (ERC Grant Agreement no.648680) and from the Academy of Finland (Grant 278722). TM is supported by the STFC grant ST/P000762/1.

\end{document}